\documentstyle[12pt,aaspp4]{article}
\lefthead{Abramenko et al.}
\righthead{Small-scale granules}

\begin{document}

\title{Detection of Small-Scale Granular Structures in the Quiet Sun with the New Solar Telescope}
\author{Abramenko, V.I.$^1$, Yurchyshyn, V.B.$^1$, Goode, P.R.$^1$, Kitiashvili, I.N.$^2$, Kosovichev, A.G.$^2$ }
\affil{
$^1$ - Big Bear Solar Observatory, 40386 N. Shore Lane, Big Bear City, CA
92314, USA;\\
$^2$ - W. W. Hansen Experimental Physics Laboratory, Stanford University, Stanford, CA 94305, USA}

\begin{abstract}
Results of a statistical analysis of solar granulation are presented. A data set of 36 images of a quiet Sun area on the solar disk center was used. The data were obtained with the 1.6~m clear aperture New Solar Telescope (NST) at Big Bear Solar Observatory (BBSO) and with a broad-band filter centered at the TiO (705.7~nm) spectral line. The very high spatial resolution of the data (diffraction limit of 77~km and pixel scale of 0.$''$0375) augmented by the very high image contrast (15.5$\pm$0.6\%) allowed us to detect for the first time a distinct subpopulation of mini-granular structures. These structures are dominant on spatial scales below 600~km. Their size is distributed as a power law with an index of -1.8 (which is close to the Kolmogorov's -5/3 law) and no predominant scale. The regular granules display a Gaussian (normal) size distribution with a mean diameter of 1050 km. Mini-granular structures contribute significantly to the total granular area. They are predominantly confined to the wide dark lanes between regular granules and often form chains and clusters, but different from magnetic bright points. A multi-fractality test reveals that the structures smaller than 600~km represent a multi-fractal, whereas on larger scales the granulation pattern shows no multi-fractality and can be considered as a Gaussian random field. The origin, properties and role of the newly discovered population of mini-granular structures in the solar magneto-convection are yet to be explored.
\end{abstract}

\section{Introduction}
 
Solar granulation is visible in broadband filter images as a cellular pattern of bright features separated by dark lanes and is regarded to be a manifestation of convection in the outermost layers of the solar convective zone (e.g., Nordlund et al. 2009). Physical properties of the granulation are relevant to energy transfer from the sub-photospheric convective layers into the photosphere. This causes perpetual research interest in solar granulation,  which elevates any time when a newer, higher resolution solar instrument comes on line (e.g., Roudier \& Muller 1986; Schrijver et al. 1997; S{\'a}nchez Cuberes et al, 2000; Danilovic et al. 2008; Yu et al. 2011). 

Roudier and Muller (1986) analyzed 0$''$.25 resolution white-light photographs of the quiet Sun granulation obtained with the Pic-du-Midi 50-cm refractor. They found a gradual decrease in granular size distribution function that implies absence of any dominant spatial scale. At the same time, these authors found that granules of about 1000 km in size contribute most to the total granular area. They referred to this 1000 km scale as a ``dominant scale'' and associated it with the characteristic thickness of the top layer of the convective zone.  Schrijver et al. (1997) utilized quiet Sun granulation images from the Royal Swedish Observatory at La Palma with an effective spatial resolution of 0$''$.4. They also reported that the granular cell size distribution is compatible with that reported by Roudier and Muller.

A key property of images of granulation is the root-mean-square of the intensity fluctuations, $\Delta I_{rms}$ (Roudier \& Muller, 1986; S{\'a}nchez Cuberes et al, 2000; Danilovic et al. 2008; Yu et al. 2011):
\begin{equation}
\Delta I_{rms}=(\Sigma(I-I_0)^2/NI_0^2)^{1/2},
\label{rms}
\end{equation}
where $I_0$ is the mean intensity of the image and $N$ is a number of data points, and the sum is taken over all image data points. Recently this parameter was referred to as the {\it granulation contrast} (see, e.g., Danilovich et al. 2008). (Note that Roudier and Muller (1986) reserved the term ``contrast'' for a different parameter.) In this paper, we will refer to  $\Delta I_{rms}$ calculated from Eq. (\ref{rms}) as the granulation contrast. 

Danilovic et al. (2008) compared the granulation contrast derived from MHD-simulations to Hinode/SP data (continuum images at 630~nm, Tsuneta et al. 2008). They reported a granulation contrast from the simulation data about 14-15\%, which was found consistent with the observed granulation contrast of 7\% , after appropriate degradation of the simulated data. According to S{\'a}nchez Cuberes et al. (2000), raw ground-based and balloon observations do not produce a granulation contrast, $\Delta I_{rms}$, higher than 9.4\%, which was obtained by Rodriguez Hidalgo et al (1992) with the 0.5~m Swedish Vacuum Solar Tower (La Palma) at a wavelength of 468.6 nm.

In this study, we present the results of analysis of the granulation contrast,  size distributions, and multi-fractal properties of solar granulation using quiet-Sun data near the disk center obtained with the 1.6 m New Solar Telescope (NST, Goode et al. 2010a,b) at the Big Bear Solar Observatory (BBSO).

\section{Data and Data Processing}

A two-hour long, uninterrupted data set was obtained with the NST on August 3, 2010 under excellent and stable seeing conditions. The solar granulation was observed in a quiet-Sun area near the disk center with a broad-band TiO filter centered at 705.7 nm with the passband of 1 nm.   This absorption line is only formed at low temperatures below 4000 K (Berdyugina et al. 2003). In quiet Sun granulation, where the temperature is higher, observations with the TiO filter register only solar continuum intensity at this wavelength (Solanki, 2011). 

The TiO images were acquired with the aid of an adaptive optics (AO) system (Denker et al. 2007, Cao et al. 2010). Only the central part (28$''$.3$\times$26$''$.3) of the entire NST field of view (77$''\times$77$''$) was utilized to take maximal advantage of the AO system.  The pixel scale of the camera, 0.${''}$0375, is 2.9 times smaller than the telescope diffraction limit of 77 km. The Kiepenheuer-Institut f{\"u}r Sonnenphysik's software package for speckle interferometry (KISIP) was utilized   to achieve the diffraction limit resolution in reconstructed images (W{\"o}ger \& von der L{\"u}he 2007). In the image reconstruction, a series of 100 images taken 12 ms apart was input to the KISIP code to produce one speckle reconstructed image.  The final data set consists of 648 speckle-reconstructed, aligned and destretched images with the time cadence of 10~s. A fragment of a granulation image is shown in Figure 1, left panel.

%#####################################################################
\begin{figure}[!ht]
\centerline{\epsfxsize=6.5truein \epsffile{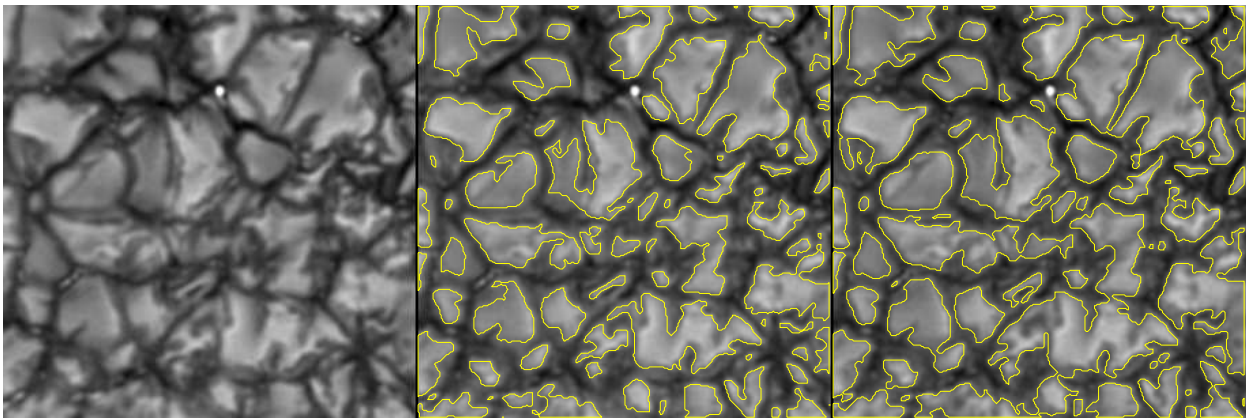}}
\caption{Left panel- A normalized image of quiet Sun granulation acquired on August 3, 2010 at 17:44:11 UT with the BBSO/NST 705~nm broadband filter imager. The pixel scale is 0$''$.0375 and the size of the FOV is 11$''$.4 $\times$ 11$''$.4 (8.2 $\times$ 8.2~Mm). The image is scaled from $0.6 I_0$ (black) to $1.6 I_0$ (white) of mean intensity, $I_0$. Middle (right) panel - The same image but overplotted with contours of $1. 03 I_0$  ($0.97 I_0$).}
\label{fig1}
\end{figure}
%#####################################################################

Each data pixel of an image was normalized by the mean intensity of that image.  A histogram of the normalized intensity calculated for all pixels of the 648 images is shown in Figure 2 (left). The histogram shows that the NST imager is able to capture a wide dynamical spectrum of intensities ranging from 0.4 to 2.5. The granulation contrast, calculated for each speckle-reconstructed image using Eq. \ref{rms} and averaged over all images, is 15.5$\pm$0.6 \%. (The average value of the granulation contrast in the central parts of the observed (raw) images is 5.1$\pm$0.07\% with the highest value of 7.1\%.)

We also calculated the granulation contrast, $\Delta I_{rms}$, from simulated intensity images obtained using the STOPRO radiative transfer code (Kitiashvili et al. 2012). For three spectral lines, which include  the blue continuum (450.45~nm), Fe~I 630.25~nm, and TiO 705.68~nm, the contrast was found to be 31.4\%, 18.1\%, and 16.1\%, respectively. We thus conclude that the contrast depends on the spectral range. After convoluting the simulated data with the NST/TiO point spread function, we found that the TiO contrast reduced to 13.6\%.

Thus the observed contrast of speckle-reconstructed images is comparable with the contrast from the aforementioned simulated data. These high contrast solar granules detected from NST/TiO images allow us to study a new level of detail in solar granulation.

%#####################################################################
\begin{figure}[!ht]
\centerline{
\epsfxsize=2.2truein \epsffile{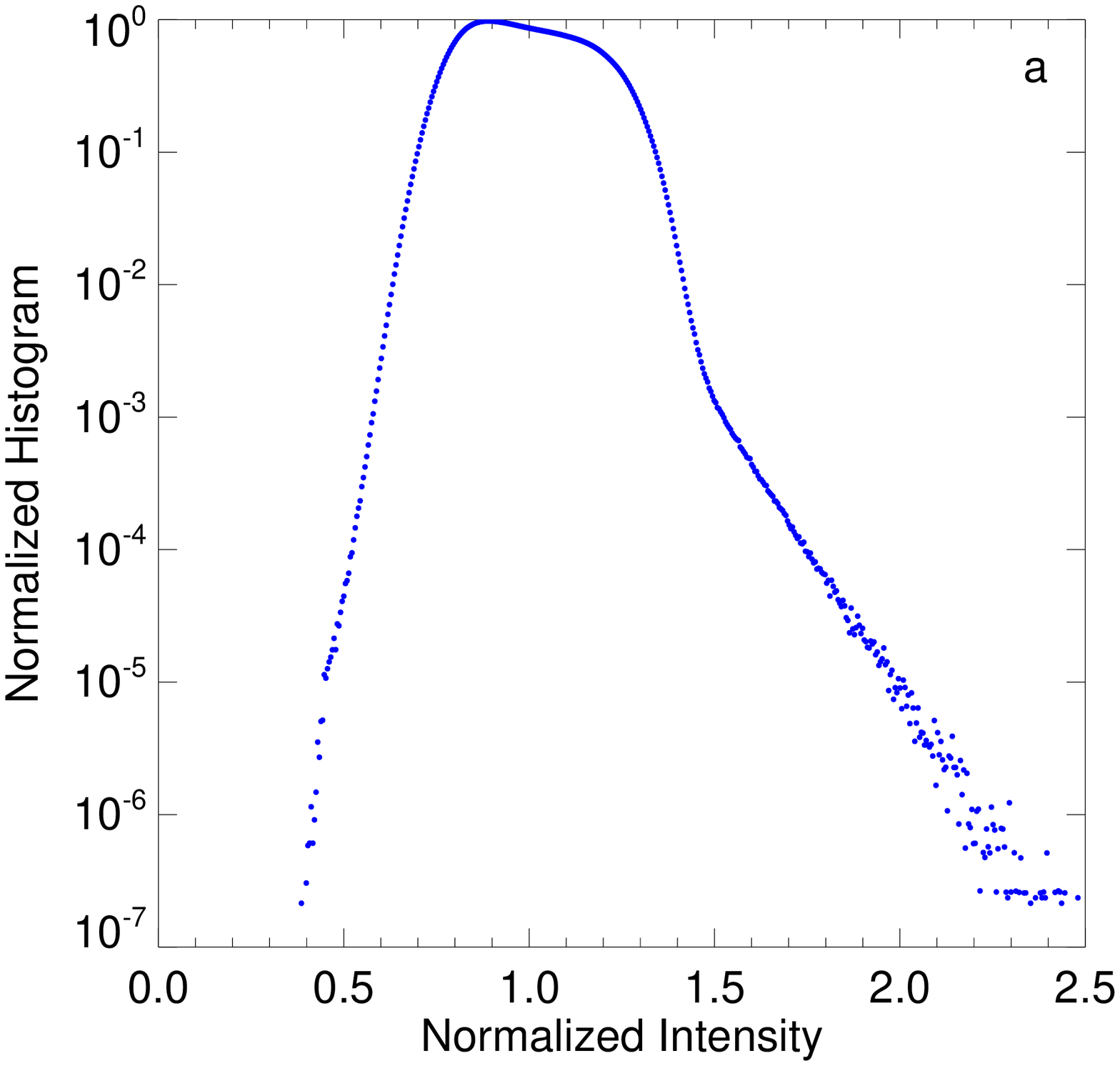}
\epsfxsize=4.5truein \epsffile{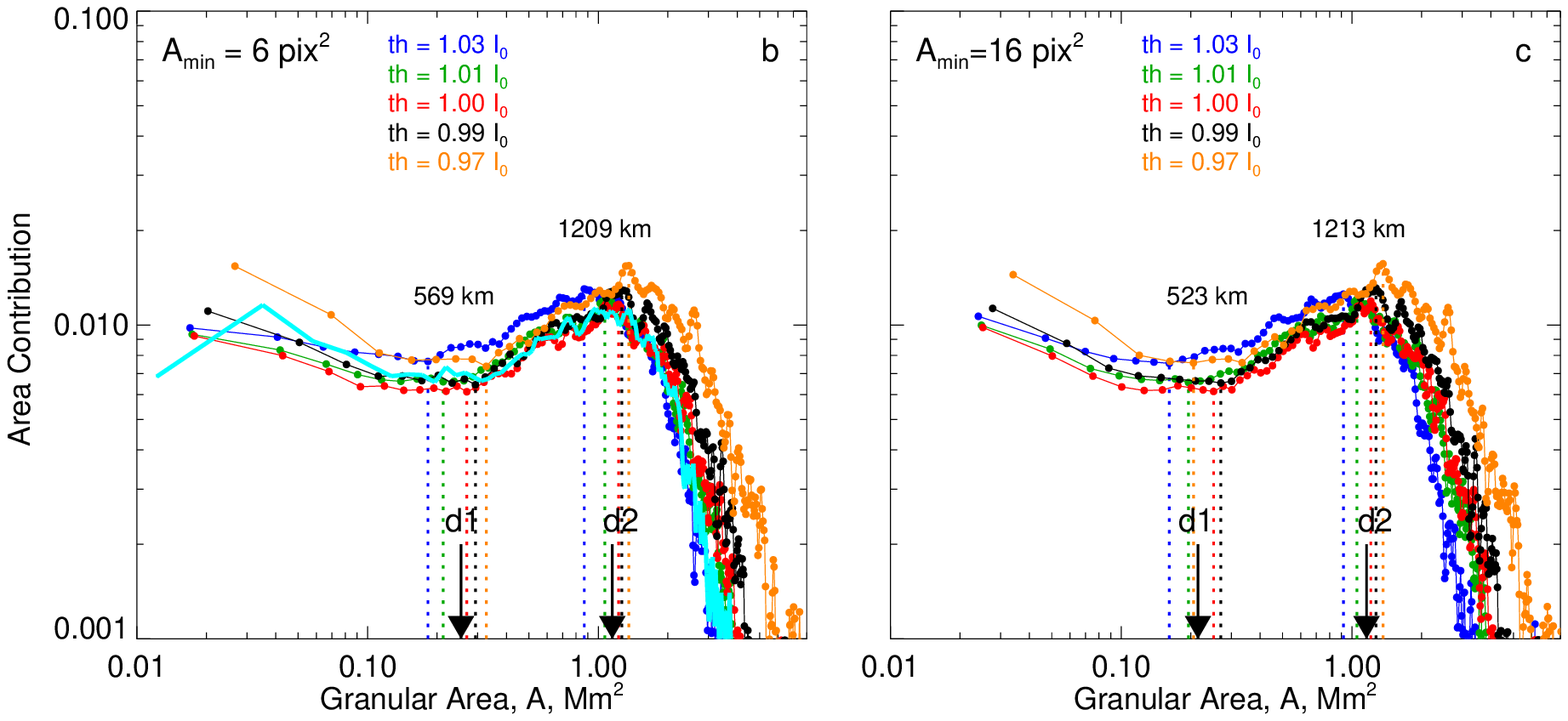}}
\caption{{\it a} - A normalized histogram of the normalized image intensity derived from the 648 images of 755$\times$700 pixels each. {\it b, c}  - Contribution of granules of a given size to the total granular area for two granule size threshold, $A_{min}$. Color curves show the data for different thresholds applied to detect granules. Granular scale d1  (d2) is determined as a local minimum (maximum) of the corresponding area contribution function. The average values of d1 and d2 are indicated in the graphs above the curves. The turquoise curve in panel {\it b} shows the result obtained from the MLT granule detection code applied for the same interval of thresholds: (1.03 - 0.97)$I_0$.}
\label{fig2}
\end{figure}
%#####################################################################

\section{Granule Detection}

For our statistical study, we selected 36 images 3 minutes apart and used a thresholding technique to detect granulation structures. We used 5 different threshold levels: 1.03, 1.01, 1.00, 0.99, and 0.97 of the mean intensity, $I_0$. Middle and right panels in Figure 1 illustrate how the threshold level affects the granule detection. With the highest threshold (middle panel) we reliably detect bright granules while missing the weak ones, or detect only their tips. In this case, the detected granules are well-separated from each other. With the lower threshold (right panel), the area of the detected granules expands, and more granules of small size and weak contrast are selected. The separation between the detected granules becomes smaller, and some of the granules merge into one entity. 

A critical parameter for granules detection is the minimum size of the detected granules. We chose to run the detection routine with the minimum size parameter set to three different levels, $A_{min}$, equal to 6, 9, and 16 squared pixels. The lowest level corresponds to the equivalent diameter of granules equal to the diffraction limit, 77 km.

To avoid a contamination of results caused by the presence of bright points (BPs), we discarded all detected entities that contained BPs. Location of BPs was determined with our BPs detection code (Abramenko et al. 2010). (Possible effects of elimination of BPs are discussed in the next section.) For each detected granule, we measured its area, $A$, and the equivalent diameter, $d$,  as the diameter of a circle of area $A$. We produced 15 detection runs (for five intensity thresholds and three minimum granule sizes) each containing about 10$^4$ granules.

Along with the traditional thresholding described above, we applied also the multiple level tracking algorithm (MLT, Bovelet \& Wiehr 2001) to detect granules. This code allows us to detect granules with regard to a hierarchy of thresholds.

\section{Two Populations of Granules}

Following Roudier \& Muller (1986), we calculated the area contribution function, which is defined as a ratio of the total area of granules of given size to the total area of all granules (Figure 2, {\it b, c}). In all 15 detection runs, the area contribution function displays a local minimum on scales of 0.16-0.31~Mm$^2$, which corresponds to equivalent diameters of 450-630 km. The averaged over the 15 runs location of the minimum is $d1=543 \pm 76$ km.  The function also shows a local maximum on scales of 0.92-1.38~Mm$^2$, which corresponds to the equivalent diameter of 1080-1320 km with the averaged value of  $1209 \pm 92$~km defined as $d2$. The existence and the location of this local maximum represents the dominant scale of granules that contribute the most to the total granular area, and it is in good agreement with the earlier results of Roudier \& Muller (1986).

However, unlike the earlier studies, our data show that the area contribution function increases on scales below $d1$. This is a previously unreported finding. The area contribution functions shown in Figure 2 suggest that the existence of the $d1$-minimum does not depend on the choice of a granule detection code. More importantly, these functions also show that small granules of sizes less than $d1$, contribute significantly to the total granular area and their contribution is comparable to that of the 1000-1300 km granules.

\begin{figure}[!ht]
\centerline{\epsfxsize=7.0truein \epsffile{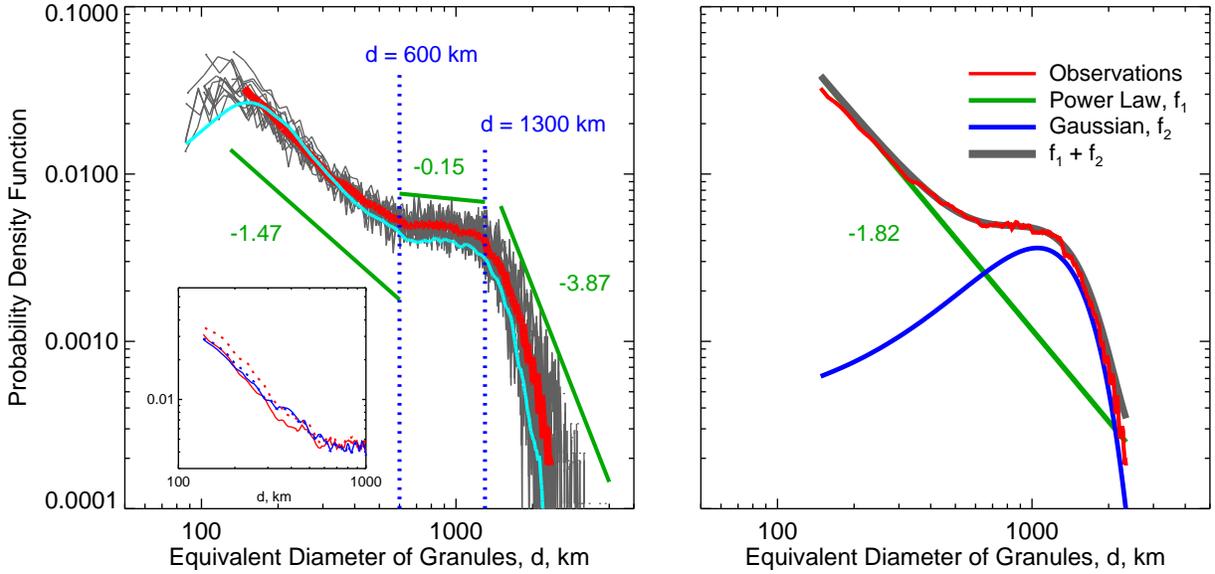}}
\caption{Left - PDFs of the granule equivalent diameter derived from 15 various detection runs (gray lines) overplotted with their average (thick red line). The turquoise curve shows the result obtained from the MLT granule detection code. Positions where the averaged PDF rapidly changes its slope are denoted by the vertical blue dotted lines. 
Right - decomposition of the observed averaged PDF (red line) into two components: a power law approximation, ($f_1$, green line) and a Gaussian approximation ($f_2$, blue). Their sum ($f_1+f_2$) is plotted with the gray line. The inset shows the PDFs calculated from the areas with low (blue) and high (red) population of BPs (the blue and red boxes in Figure \ref{fig4}, correspondingly); solid (dashed) lines are the PDFs calculated after (before) elimination of BPs. 
}
\label{fig3}
\end{figure}

Figure \ref{fig3} shows the probability density functions (PDFs) of the equivalent granular diameter, $d$. The left graph shows the PDFs plotted for all 15 detection runs (gray lines). The result from the MLT code applied for the same interval of thresholds and $A_{min}=$6 pixels is also shown for comparison.
%#####################################################################
%#####################################################################
Signatures of the distinct scales $d1$ and $d2$ discussed above are clearly visible in Figure 3, as well. On scales of approximately 600 and 1300 km, the averaged PDF rapidly changes its slope. This varying power law PDF is suggestive that the observed ensemble of granules may consist of two populations with distinct properties. We thus attempted a two component decomposition of the observed PDF as a combination of a power law function, $f_1$, and a Gaussian function, $f_2$ (Figure 3, right):
\begin{equation}
f_1=C_1 d^{\kappa}
\label{PL}
\end{equation}
\begin{equation}
f_2=\frac{C_2}{\sigma(2\pi)^{1/2}}exp \left (-\frac{1}{2} \left (\frac{d-d_0}{\sigma} \right )^2 \right ).
\label{Gaus}
\end{equation}
Here, $d_0$ and $\sigma$  are the mean and the standard deviation of $d$. The parameters of the best fits are: $C_1=10^{2.53}$ km$^{-\kappa}$, $\kappa=-1.82\pm 0.12$ (dimentionless coefficient), $C_2=4.35 \pm 0.09$ km$^{-1}$, $\sigma=480 \pm 11$ km, and $d_0=1050 \pm 22$ km with the reduced $\chi^2-$value of 0.38.

The successful decomposition suggests that the entire ensemble of granules can be considered to be a co-existence of two distinct populations. Regular granules of a typical size of 600-1500~km across constitute the Gaussian subset.  Along with them, there apparently exists a subset of granular structures with the power-law distribution across all scales, from 130 to 2000 km. On scales smaller than approximately 600~km, this population becomes dominant, and we call these structures mini-granules.

%#####################################################################
\begin{figure}[!ht]
\centerline{\epsfxsize=5.0truein \epsffile{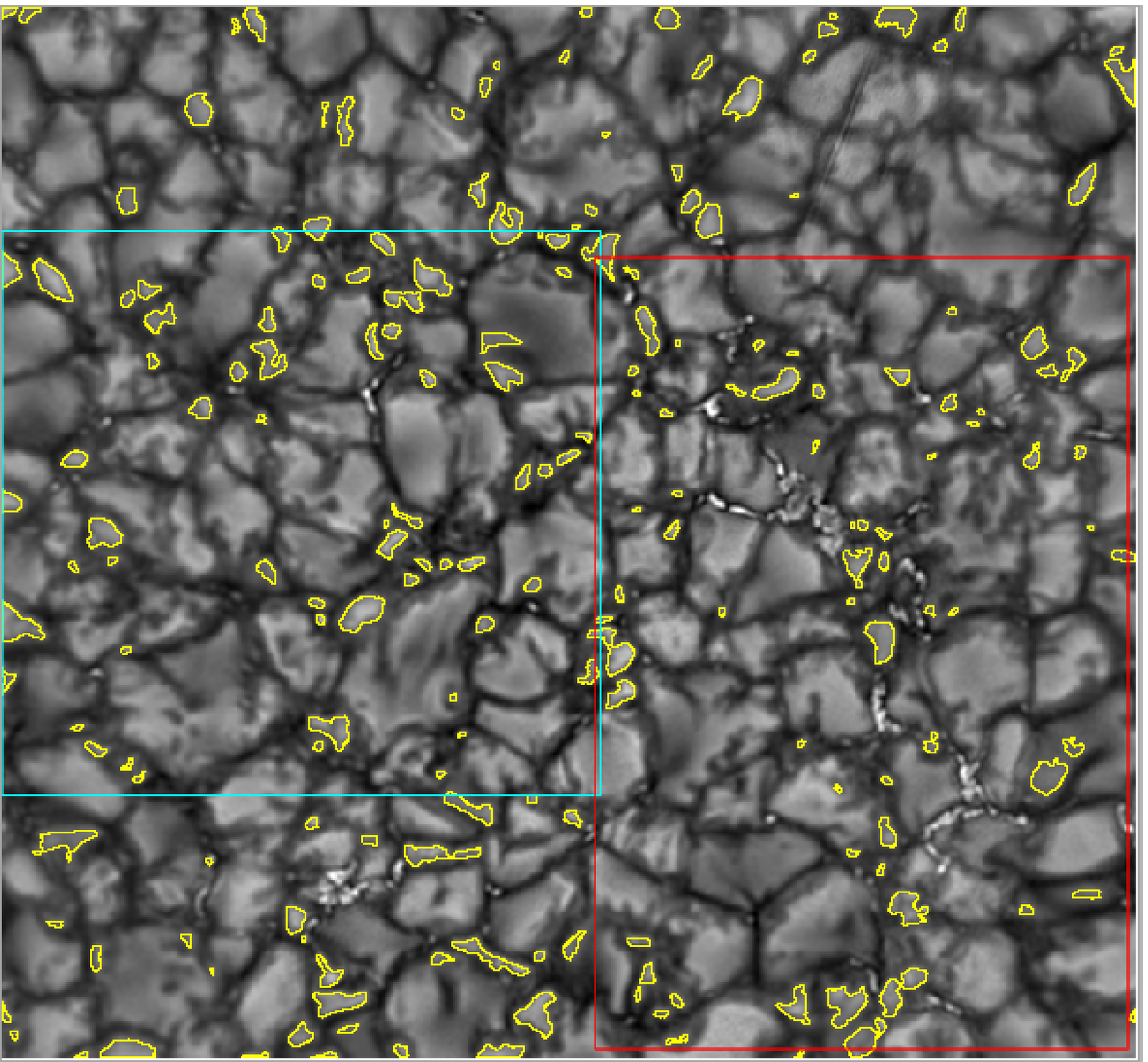}}
\caption{Image of solar granulation acquired at 18:30:30~UT on August 3, 2010 with overplotted contours of detected mini-granules (\textit{i.e.}, structures of equivalent diameter less than 600 km). This particular plot was obtained with a threshold of 1.0$I_0$. The background image is scales from 0.6$I_0$ (black) to 1.6$I_0$ (white).The blue and red boxes outline areas of low and high population of BPs, respectively. The image size is 755$\times$700 pixels, or 28$''$.3$\times$26$''$.25. The full resolution image can be found at  http://www.bbso.njit.edu/$\sim$avi/ } 
\label{fig4}
\end{figure}
%#####################################################################

The mini-granular structures appear to be confined to broad inter-granular lanes (Figure 4), frequently forming chains and clusters, in contrast to the regular granules, which are distributed more evenly over the solar surface. Note, that there are uncontoured (undetected)  mini-granular structures in the image. These were missed by the detection routine either because of their below the threshold intensity or because they were contoured together with the larger neighboring granule. Nevertheless, it is evident that the mini-granules are neither regular granules, nor magnetic BPs (recall that BPs were eliminated from consideration).

To clarify how BPs can affect the calculated PDFs, we undertook the following experiment. We selected two areas inside the FOV with different surface density of BPs. The blue box in Figure 4 encloses an area with almost no BPs, while the red box includes clusters of BPs. For each area, we calculated two PDFs: one without elimination of BPs (dotted lines in the inset in Figure \ref{fig3}) while another PDF was derived after elimination of BPs (solid lines in the inset in Figure \ref{fig3}). Elimination of BPs does not affect the PDF calculated from the blue box area (very few BPs). In the red box (clusters of BPs), influence of BPs is only detectable as a slight enhancement of the  probability density on small scales below 300 km. After elimination of BPs, the red-box PDF follows the rest of the PDFs. This experiment demonstrated that the possible contamination of the PDFs by residual BPs is negligible.

The intermittent spatial distribution of mini-granular structures was further studied by probing the multi-fractal properties of the images of granulation. We utilized the structure function method (Abramenko et al. 2002) that allows us to compute the flatness function, $F(r)$, and thus characterize the degree of multi-fractality and intermittency on different linear scales (Abramenko 2005; Abramenko \& Yurchyshyn 2010). Calculations of the flatness function do not involve thresholding and do not have free parameters. Thus, this approach is an independent means of estimating characteristic scales of solar granulation. 

%#####################################################################
\begin{figure}[!ht]
\centerline{\epsfxsize=4.0truein \epsffile{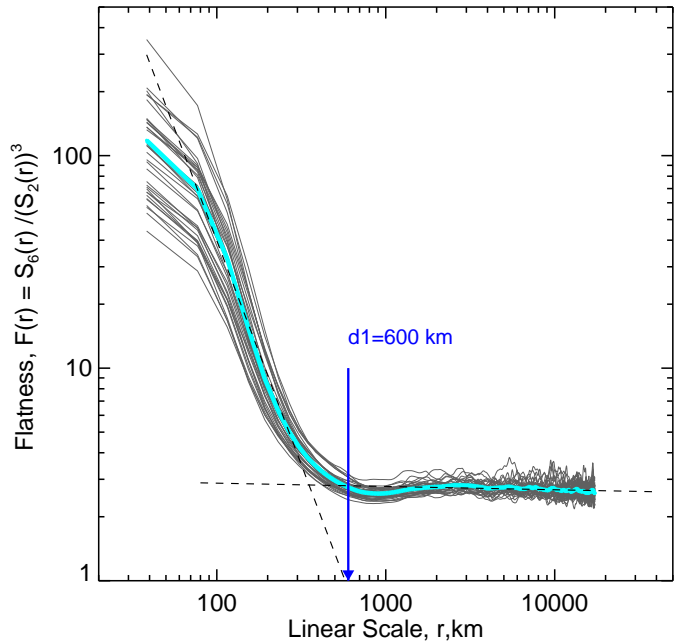}}
\caption{Flatness functions calculated from 36 granulation images (gray) and their average (turquoise). The blue arrow divides the multi-fractality range where the flatness function varies as a power law from the Gaussian range where the flatness function is scale independent.} 
\label{fig5}
\end{figure}
%#####################################################################

We calculated $F(r)$ as a ratio of the 6th-order structure function, $S_6(r)$, to the cube of the second-order structure function, $S_2(r)$. A $q-$order structure function is defined as the $q$-th power of an increment of intensity in two pixels separated by a distance $r$, then averaged over the entire image. The flatness functions calculated for the 36 granulation images are plotted with gray lines in Figure 5. 
Generally speaking, in the case of highly intermittent and multi-fractal nature of features inside
a certain scale range, the flatness function behaves as a power law with a negative exponent (see, e.g., Frisch 1995, Abramenko 2005). Conversely, a Gaussian process (no intermittency, no multi-fractality) is associated with a ``flat'' flatness function, i.e., it does not depend on spatial scales (see, \textit{e.g.}, Abramenko 2005). Figure 5 thus clearly demonstrates that regular granules are not associated with any intermittent (or, in other words, multi-fractal) process and therefore one expects a more or less random (Gaussian) distribution of dark and bright features over the surface on scales above 600~km. At the same time, the mini-granular structures ($d<600$~km) are distributed over the surface in a highly irregular, intermittent manner forming a multi-fractal pattern (when bunches of elements are intermittent with voids).

We thus further confirmed that the appearance of the characteristic scale of $d1=600$ km is not related to thresholding of images and represent real change in properties of solar granulation. The flat segment of $F(r)$ function on scales larger than 600 km also confirms our choice of the Gaussian function as the best fit for regular granule size distribution (Figure 3, right panel). The power-law behavior of $F(r)$ function on small scales ($<600$ km) is compatible with a power law distribution of mini-granules sizes: both of them are manifestation of intermittent and multi-fractal nature of the measured variable (see, e.g., Schroeder 2000, Abramenko 2008, Aschwanden 2011 and references herein).

\section{Conclusions}

Solar granulation data measured with the NST's broad-band TiO imager display a very high granulation contrast (r.m.s. brightness fluctuation, in the terminology of Roudier \& Muller, 1986) of 15.5$\pm$0.6\%, which agrees very well with the the 14-16\% contrast for the TiO spectral line inferred from the numerical simulations using the STOPRO radiative transfer code (Kitiashvili et al. 2012). The NST contrast also comparable with the 14-15\% contrast inferred from 3D MHD model data for the Hinode 630.25~nm spectral line (Danilovic et al. 2008). This high-contract and high spatial resolution data from the NST (diffraction limit of 77 km, pixel scale of 0$''$.0375) allowed us to perform a statistical study of the granule size distribution and explore characteristic scales of solar granulation in detail.
The results formulated below do not depend on the thresholding technique: The traditional single-threshold technique and the MLT technique produced similar results.

Analyzing 36 independent granulation images, we concluded that there are two populations of granular structures. First population is comprised of 
relatively large (regular) granules of the mean size of approximately 1000~km across. The second  a population includes "mini-granules" - a continuous power-law
population of convective structures that dominate on scales less than approximately 600~km.

The size histogram of regular granules is approximated by a normal (Gaussian) distribution function with a mode of 1050~km and the standard deviation of 480~km. Granules with sizes of 1080-1320~km across (with a mean value of about 1200~km) contribute substantially to the total area of granules (see Figure 2). These regular granules appear to be randomly distributed over the solar surface. We further confirm existence of the characteristic (or ``dominant'' per Roudier \& Muller, 1986) scale of granules. We found it to be about 1080-1300~km (1$''$.49 - 1$''$.79) from the area contribution function (Figure 2) and 1050$\pm$480~km from the PDF approximation (Figure 3).

The size distribution of mini-granules can be approximated with a power-law function with a slope of -1.82$\pm$0.12. Their contribution to the total granular area is comparable to that of regular granules. The mini-granules are mainly confined to broad inter-granular lanes and form chains and clusters intermittent with voids.  

The flatness function (a measure of intermittency and multi-fractality) calculated directly (without any thresholding) from the images indicates that solar granulation is non-intermittent on scales exceeding 600 km, and it becomes highly intermittent and multi-fractal on smaller scales. Thus, a random Gaussian-like distribution of granules over the solar surface holds down to 600 km only. On smaller scales, the multi-fractal spatial organization of the mini-granular structures takes over.

Using Hinode data, Yu et al. (2011) reported two types of granules: small and large ones. The dividing point between them is 1044~km, which exceeds significantly that found in our study (600~km). The 1044~km scale is very close to the mode (1050~km) of the Gaussian distribution of the regular granules studied here. More importantly, the minimum size of granules studied in Yu et al. (2011) is half of an arc-second (360~km). In other words, the small granules reported by these authors correspond mainly to the plateau in the size distribution reported here (see Fig. 3), while the subset of mini-granules was mostly missed in their study.

A possible interpretation of mini-granular structures is that they are fragments of regular granules, which are subject to highly turbulent plasma flows in the intergranular lanes, where the intensity of turbulence is enhanced (Nordlund et al. 2009). The association between the mini-granulation and magnetic and velocity fields, as well as efforts to detect mini-granulation in numerical simulations of solar magneto-convection are subjects for future research.  As for now, it is evident that the complex picture of solar near-surface magneto-convection might be even more complex.

We are thankful to anonymous referees whose comments helped us to improve the paper.
Authors gratefully acknowledge help of the NST team and support of NSF (ATM-0716512 and ATM-0847126), NASA (NNX08AJ20G, NNX08AQ89G, NNX08BA22G), AFOSR (FA9550-12-1-0066). 

{}
\end{document}